\documentclass[]{spie}  

\usepackage{amsmath,amsfonts,amssymb}
\usepackage{graphicx}
\usepackage{enumitem}
\usepackage{booktabs}
\usepackage{enumitem}

\usepackage[colorlinks=true, allcolors=blue]{hyperref}

\title{The sky at one terabit per second: Architecture and implementation of the Argus Array Hierarchical Data Processing System}

\author[a]{Hank Corbett}
\author[a]{Alan Vasquez Soto}
\author[a]{Lawrence Machia}
\author[a]{Nathan Galliher}
\author[a]{Ramses Gonzalez}
\author[a]{Nicholas M. Law}
\affil[a]{Department of Physics and Astronomy, University of North Carolina at Chapel Hill, Chapel Hill, NC 27599-3255, USA}

\authorinfo{Further author information: (Send correspondence to H. Corbett)\\H. Corbett: E-mail: htc@unc.edu\\}

\begin{document} 
\maketitle

\begin{abstract}
The Argus Optical Array is a synoptic survey observatory, currently in
development, that will have a total collecting area equivalent to a 5-meter
monolithic telescope and an all-sky field of view, multiplexed from 900
commercial off-the-shelf telescopes.
The Array will observe 7916 deg$^2$ every second during high-speed operations
($m_g\leq16.1$) and every 30 seconds at base cadence ($m_g\leq19.1$),
producing 4.3 PB and 145 TB respectively of data per night with its 55-gigapixel mosaic of cameras.
The Argus Array Hierarchical Data Processing System (Argus-HDPS) is the
instrument control and analysis pipeline for the Argus Array project, able to create
fully-reduced data products in real time. 
We pair sub-arrays of cameras with co-located compute nodes, responsible for
distilling the raw 11 Tbps data rate into transient alerts, full-resolution
image segments around selected targets at 30-second cadence, and full-resolution
coadds of the entire field of view at $15+$-min cadences. 
Production of long-term light curves and transient discovery in deep coadds out
to 5-day cadence ($m_g\leq24.0$) will be scheduled for daytime operations. 
In this paper, we describe the data reduction strategy for the Argus Optical
Array and demonstrate image segmentation, coaddition, and difference image
analysis using the GPU-enabled Argus-HDPS pipelines on representative data from
the Argus Array Technology Demonstrator.

\end{abstract}


\section{The Argus Optical Array: Phased Prototyping of a deep multiplexed survey}\label{sec:intro} 

The Argus Optical Array will be a massively-multiplexed optical observatory,
designed for synoptic time-domain observations. We are currently undertaking a
phased prototyping process, evaluating hardware for an eventual 900-telescope
array which would provide comparable collecting area to a 5-meter monolithic
mirror telescope. By using a large array of 200-mm, commercially-available
telescopes, this collecting area can be spread out over an 8000 sq. degree field
of view, enabling simultaneous monitoring of O($10^7$) stars and galaxies. The
full survey description and science justification for the Argus Optical Array is given in
Ref.~\citenum{2022PASP..134c5003L}, and an updated mechanical and optical design can be
found elsewhere in the Proceedings.\cite{argus_spie} A full treatment of the
general utility of multiplexed array observatories relative to monolithic
systems for high-speed sky surveys can be found in
Ref.~\citenum{2020PASP..132l5004O}.

The first stage in this process, the 9-telescope Argus Array Technology
Demonstrator (A2TD) based on the design concept of the
Evryscopes,\cite{evryscope_instrument,evryscope_project} was completed in 2021.
A2TD is described in full elsewhere in the Proceedings.\cite{corbett_spie2022b}
A2TD provides a lab-local test bed for rapid prototyping of control systems,
data management, and other enabling technologies for Argus Array array,
including a novel automated polar alignment and single-axis tracking
drive,\cite{motion_control} on-sky testing of climate control systems,
\cite{hvac_argus} mechanical and structural support,\cite{hercules_mount}.

The second stage of development, Argus Pathfinder, will contain 38 telescopes,
enabling it to survey the seasonal northern sky between declinations of
$-20^\circ$ and $72^\circ$ every night. Pathfinder will observe at 30-second
cadence, with a $5-\sigma$ limit of $m_A=19.6$ in single images, where $m_A$
is the magnitude in a wide passband ranging from 350 nm to the blue edge of the
Fraunhofer A line and O$^2$ telluric band at 750 nm. This depth is equivalent to
$m_g=19.1$ or $m_r=18.3$ under dark sky conditions. Argus Pathfinder will be the
first iteration to be deployed to a dark site, the Pisgah Astronomical Research
Institute in Rosman, NC.

The Argus Array observing strategy, used at all stages of instrument
development, uses a ratcheting approach to build up sky coverage over
the course of a night. The instrument tracks the sky in 15
minute intervals, each of which concludes with a rapid slew back to the next
field position, defined by the pointing of the meridian of the array. Two
observing cadences will be used; a standard survey with 30-second
exposures, and a secondary fast survey at 1-second cadence. In both cases, the
dead time between consecutive images is sub-millisecond, with images read
out to an internal camera buffer before transfer to a control computer over
USB3.0.

Table~\ref{tbl:array_facts} contains the system parameters of Argus
Pathfinder and the Argus Optical Array instruments, including properties of the
dataset.
\begin{table}[ht]
    \begin{center}       
    \begin{tabular}{c | c c}
    {} & \textbf{Argus Pathfinder (Q3 2022)} & \textbf{Argus Optical Array} \\ 
    \hline
    \rule[-1ex]{0pt}{3.5ex} {Telescopes} &  38x Planewave 203 mm F/2.8  &     900 x Planewave
    203 mm F/2.8 \\ 
    {} & (combined 1m-class equiv.) &   (combined 5m-class equiv.) \\ 
    \rule[-1ex]{0pt}{3.5ex} \textbf{Detectors} & 61 MPix Sony IMX455 sCMOS  & 61 MPix Sony IMX455 sCMOS  \\ 
    {} &  1.7e- RN and 80 $\mu$s dead time & 1.7e- RN and 80 $\mu$s dead time    \\
    {} &  $>90\%$ QE at 475 nm & $>90\%$ QE at 475 nm \\ 
    \rule[-1ex]{0pt}{3.5ex} \textbf{Field of View} &  9 sq. deg per telescope  &  9 sq. deg per telescope \\ 
    {} &  344 sq. deg instantaneous & 7916 sq. deg instantaneous \\ 
    \rule[-1ex]{0pt}{3.5ex} \textbf{Nightly Sky Coverage} &  19,370 sq. deg.  &  19,370 sq. deg. \\ 
    {} &  (15 minutes per night) & (2-10 hours per night) \\ 
    \rule[-1ex]{0pt}{3.5ex} \textbf{Pixel Sampling} &  1.38 arcsec/pix  &  1.38 arcsec/pix \\ 
    \rule[-1ex]{0pt}{3.5ex} \textbf{Site} &  North America  & North America \\ 
    {} &  Pisgah Astronomical & Pisgah Astronomica \\ 
    {} &  Research Institute  & Research Institute\\ 
    \rule[-1ex]{0pt}{3.5ex} \textbf{Exposure Time} &  1 second high-speed & 1-second high-speed \\ 
    {} &  30-second base-cadence  & 30-second base-cadence \\ 
    \rule[-1ex]{0pt}{3.5ex} \textbf{Wavelength} & Wide-band (350-750 nm) & Wide-band (350-750 nm) \\ 
    {} &   & or alternating g', r' \\ 
    \rule[-1ex]{0pt}{3.5ex} \textbf{Pixel Count} & 2.3 GPix & 54.9 GPix\\ 
    \rule[-1ex]{0pt}{3.5ex} \textbf{Mosaic Image Size} & 4.7 GB & 110.1 GB\\ 
    \rule[-1ex]{0pt}{3.5ex} \textbf{Nightly Raw Data} & 180 TB (high-speed) & 4.3 PB (high-speed)\\ 
    {} & 6 TB (base-cadence) & 145 TB (base-cadence)\\ 
    \rule[-1ex]{0pt}{3.5ex} \textbf{Throughput} & 464 Gbps (high-speed) & 11 Tbps (high-speed)\\ 
    \textbf{at 92\% duty cycle} & 15.5 Gbps (base-cadence) & 367 Gbps (base-cadence)\\ 
 \end{tabular}
 \end{center}
 \caption{\label{tbl:array_facts}Survey and dataset parameters for the Argus
 Pathfinder and Argus Optical Array instruments. Data types assume 16-bit pixel
 data.}

 \end{table} 

With its 54.9 GPix combined mosaic imager, the Argus Optical Array will produce
4.3 PB of raw data per night when observing in the 1-second cadence mode, and 145 TB per
night at base cadence, numbers that are comparable to the entire data sets of
the many data-rich astronomical surveys currently operating. The prototype Argus
Pathfinder system, containing 38 telescopes, will produce accordingly less
data, but still up to 180 TB per night at 1-second cadence, 24\% more than the
full Array at 30-second cadence. To reduce this data into an event stream
of astrophysical transients, long-term lightcurves, and image data products
requires both a physical compute architecture capable of the necessary throughput
and a performant software pipeline. 
 
In this paper, we introduce the Argus Hierarchical Data Production System
(Argus-HDPS), the unified control and analysis pipeline for the Argus Optical
Array project. In Sections~\ref{sec:requirements} and \ref{sec:engineering_goals}, we describe the science and
engineering requirements of the system In
Section~\ref{sec:system_archictecture}, we outline the data flow and physical
compute hardware underlying Argus-HDPS. In
Section~\ref{sec:implementation}, we describe the algorithms and
implementation status of the pipeline. Finally, in Section~\ref{sec:summary}, we
summarize and present pipeline performance results on data from A2TD.

\section{Science Requirements and Data Products}\label{sec:requirements}

Transient alerts and images are produced at multiple effective cadences, using
coaddition to perform deep searches for slow-rising transients at timescales out
to 5 days. To minimize data backlog and support rapid community followup of
transient events, HDPS must produce alerts and reduced images within cadence for
cadences less than 1 day; i.e., 30-second images must be reduced and alerts
generated within 30 seconds. As the data rate slows for longer, multi-day
coadds, sub-cadence latencies can be achieved using daytime operations.

Full-frame images from the full Argus Optical Array will be prohibitively large
to store, requiring up to 145 terabytes per night at 30-second cadence and 4.3
petabytes per night at 1-second cadence. To support long-term retention of image
data, HDPS must reduce incoming images into pre-defined data products, including: 
\begin{enumerate}
    \item Images, segmented into 13.7$\times$13.7 arcminute sky regions. 
        \begin{enumerate}
            \item Full-resolution segments (HEALPix\cite{2005ApJ...622..759G} NSIDE=256)
            at base cadence, cached locally for at least 5 days
            \item Full-resolution segments, coadded at 15-minute to 5-day cadences
            \item Sparse full-resolution segments, containing transient detections and
            pre-selected science targets 
            \item Low-resolution segments with 10$\times$ reduced resolution (13.8 arcsec/pixel). 
        \end{enumerate}
    \item Transient alerts, distributed via community brokers
    \begin{enumerate}
        \item From single images, at both 1- and 30-second cadence
        \item In deep coadds, up to 5 days
    \end{enumerate}
    \item Photometric light curves 
    \begin{enumerate}
        \item Transient sources: from image subtraction, sequentially released
        via alerts
        \item Detrended long-term lightcurves for an input catalog of O($10^7$)
        sources
    \end{enumerate}
\end{enumerate}

Reduced image data from Argus, including deep coadds, sparse full-resolution
images covering science targets, and low-resolution 13.8-arcsecond per pixel
images of the entire sky, will be stored long-term and made available publicly."

The production of long-term photometric light curves, containing tens of trillions
of photometric measurements across a five year survey, is a unique challenge among the
Argus Optical Array data products. Removing systematics from the lightcurves
inherently depends on long-term trends in time, and for this reason, we elect to
produce precision lightcurves in periodic data releases, rather than within the
observing cadence. Per-epoch photometry, however, will be produced in real time, using the
pipelines we have developed for the Evryscopes.\cite{evryscope_instrument}
Calibrated and detrended lightcurves for pre-selected science targets will be 
generated and released on a schedule.

\section{Engineering Goals}\label{sec:engineering_goals}

Analysis and control pipelines are a long-term investment in the project, and
scalability in software instrumentation will enable rapid iteration of the
physical instrument and continuity towards the full Argus Optical Array. In
addition to the science-driven goals above, we are developing Argus-HDPS with the
following engineering goals: 
\begin{enumerate}[topsep=0pt,itemsep=-1ex,partopsep=1ex,parsep=1ex]
    \item Scalability and adaptability from the prototype stage to the 900-telescope Argus Optical Array
    \item Integration between the control system and analysis system to minimize
    latency and intermediate storage requirements
    \item Maintainability of the codebase, using standard version control,\footnote{\url{https://git-scm.com}}
    dependency management,\footnote{\url{https://python-poetry.org}} and automated testing\footnote{\url{https://pytest.org}} tools
    \item Reliability for long-term operation and production of a stable
    dataset, including hardware-in-the-loop testing of cameras and focusers
\end{enumerate}
Particularly at the current stage, adaptability to a heterogeneous instrument has
been essential to enable rapid prototyping while navigating the supply chain
constraints early in the prototyping process, and will remain an important
feature of HDPS to support long-term evaluation of telescopes and cameras with
Argus Pathfinder.
Between the A2TD and Pathfinder instruments, Argus-HDPS includes support for two
camera manufacturers (Atik Cameras\footnote{\url{https://www.atik-cameras.com}}
and QHYCCD\footnote{\url{https://www.qhyccd.com}}), five camera models, and two
telescopes with associated focusing hardware (the
Celestron\footnote{\url{https://www.celestron.com}} RASA-8 and a custom
Astrograph from Planewave Instruments\footnote{\url{https://planewave.com}})

\section{System Architecture}\label{sec:system_archictecture} 

\begin{figure}
    \begin{center}
    \begin{tabular}{c}
    \includegraphics[width=0.7\columnwidth]{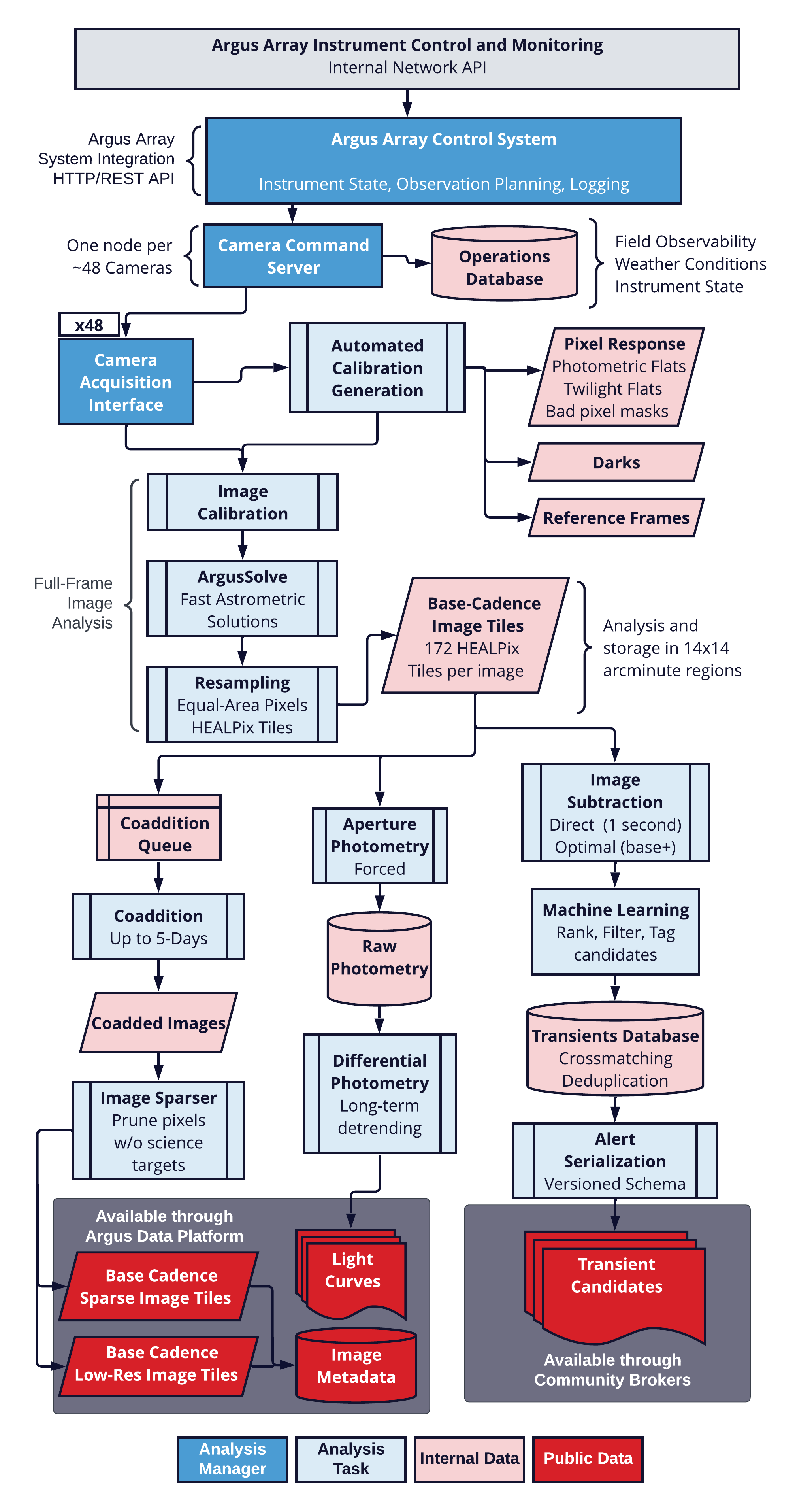}
    \end{tabular}
    \end{center}
    \caption{\label{fig:pipeline_flow} Pipeline components, processes, and data
    products for the Argus Optical Array.}
 \end{figure}

To meet the science and engineering goals of the Argus Optical Array, while
coping with data rates into the terabit-per-second regime, we use a converged
data collection and analysis pattern, in which images are reduced into data
products within the observing cadence. In Figure~\ref{fig:pipeline_flow}, we
present a flowchart of the basic components of Argus-HDPS, including the core
processes and data products. 

Integration between hardware control and data analysis systems is essential for
reliable data caching and storage management, and for latency optimization at
the fastest cadence. HDPS uses a modular software architecture for hardware
control, based on a stable contract between core system components (optical and
mechanical control, weather and instrument state monitoring, and pipeline
instances), defined using HTTP APIs. 

\subsection{Camera Control}

Each camera is directly connected to a camera command (CC) compute node capable
of reducing its data in real time. We have developed Argus-HDPS wrappers for the
vendor-supplied camera SDKs from Atik Cameras and QHYCCD, which abstract the
basic functions of the cameras to a standard, shared Python API. Each CC node
communicates with state machine processes exporting this API using an internal
socket server. To minimize latency, camera control processes copy the image
data, along with corresponding metadata, directly into a shared memory object
store (Apache Plasma\footnote{\url{https://arrow.apache.org}}), from which
images are written to disk on a rolling basis. Cameras attached to each CC node
can be controlled collectively using an asynchronous HTTP API by the instrument
control system. The control system can control an arbitrary number of CC nodes
(19 for the full Argus Optical Array).

\subsection{Compute Nodes}

CC nodes are standard many-core x86 rack-mounted servers, co-located with the
instrument. Individual cameras connect to the CC nodes via USB 3.0 and
high-density PCIe expansion cards. CC nodes will each be equipped with a GPU for
image analysis. GPU requirements are primarily driven by memory size and bandwidth; in
addition to the images themselves, each camera has a corresponding set of
calibration frames (darks, flats, bad pixel masks, plus intermediate background
and noise maps) and reference frames, which must also be transferred to the GPU
at the start of each ratchet. Each compute node is also equipped with $\sim$50
TB of local storage for temporary caching of full-frame data at 30-second
cadence, and enough RAM to cache data at 1-second cadence for up to  one minute. 

\section{Pipeline Implementation}\label{sec:implementation}

Once each full-frame image has been recorded in the in-memory object cache,
analysis is delegated to analysis processes. Each analysis process
pre-allocates the relevant reference and calibration frames on the CC node's GPU
resources at the start of each pointing. CMOS image calibration, source
detection, and resampling to the HEALPix grid are done directly on the GPU, and
source catalogs are transferred to system memory for crossmatching and
astrometric fitting on the CPU. In the current implementation, image
subtraction and coaddition are also done on the CPU. 

\subsection{Image Management}\label{sec:image_data}

The highest-data-rate components of Argus-HDPS are those which interface with
with the full-resolution 61-MPix images from the cameras. These images are both
unwieldy to access, with 122 MB file sizes that incur a disk read penalty on the
order of the exposure time, and require external indexing via a database to be
searchable in space or time. To support these standard usage patterns, we have
developed a hierarchical, equal-area storage system for Argus Array data,
inspired by, but distinct from, the Hierarchical Progressive Survey (HiPS)
format described in Ref.~\citenum{2017ivoa.spec.0519F}. 

In the full-frame images, we perform standard image calibrations (dark
subtraction, flat fielding, masking of bad pixels) and fit an astrometric
solution using a custom high-speed solver. The ArgusSolve astrometry algorithm
is based on the quadrilateral hashing~\cite{astrometry_net} and the iterative
fitting.\cite{2019PASP..131e4504O} Combined with an optimized implementation in
Python, a one-directional mapping from celestial to pixel-plane coordinates for
full-frame images can be produced in $\sim100$ milliseconds. A full
WCSLIB\cite{2011ascl.soft08003C}-compatible FITS header can optionally be
generated for the full-frame image in $\sim5$ s. 

\subsubsection{Image Segmentation}

Each full-frame image is reprojected into equal-sky-area tiles (based on the
HEALPix\cite{2005ApJ...622..759G} pixelization scheme with NSIDE=256), providing
786,432 potential 13.7$\times$13.7 arcminute sky regions across the celestial
sphere, of which 58.6\% will be observable by a Northern Hemisphere Argus
Optical Array. The resulting segment images are stored in a structured tree of
directories, grouped based on membership of each segment in the hierarchy of
HEALPix tiles with NSIDE=4,16,64. Grouping the directories in this way is done
to minimize the number of files per directory. Within each directory, images are
stored per-night as multi-extension cubes in a FITS-like format modified to
support 16-bit float data. Given a time and sky position, this scheme allows the
path for a given spatial/temporal image tile to be uniquely determined without
an external index, while maintaining a number of files on disk that is
manageable with standard filesystems. 

\subsubsection{Sparse Image Segments}

Each set of image segments for a single full-frame image is approximately the
same size as the original data; however, each image tile can be further
partitioned and selectively compressed. We have developed a ``stamping''
procedure, in which ``minipix'' segments of each image tile (at a higher HEALPix
level, corresponding to NSIDE=16384) not containing science targets (either
those detected in image subtractions or pre-cataloged by the science team) are
set to zero and the resulting image compressed. To minimize the impacts on
serendipitous science cases, full resolution images at base cadence will be
cached at least up to the longest coaddition epoch (5 days), and low resolution
maps, at a pixel scale of 13.8 arcsec per pixel, will be saved for the whole
sky. The compressed sparse and low-resolution segment images combined take up $\sim5\%$ the
storage space of the full-resolution segments.

\subsubsection{Deep Coaddition in Sky Segments}

\begin{figure}[ht]
    \begin{center}
    \begin{tabular}{c}
    \includegraphics[width=0.8\columnwidth]{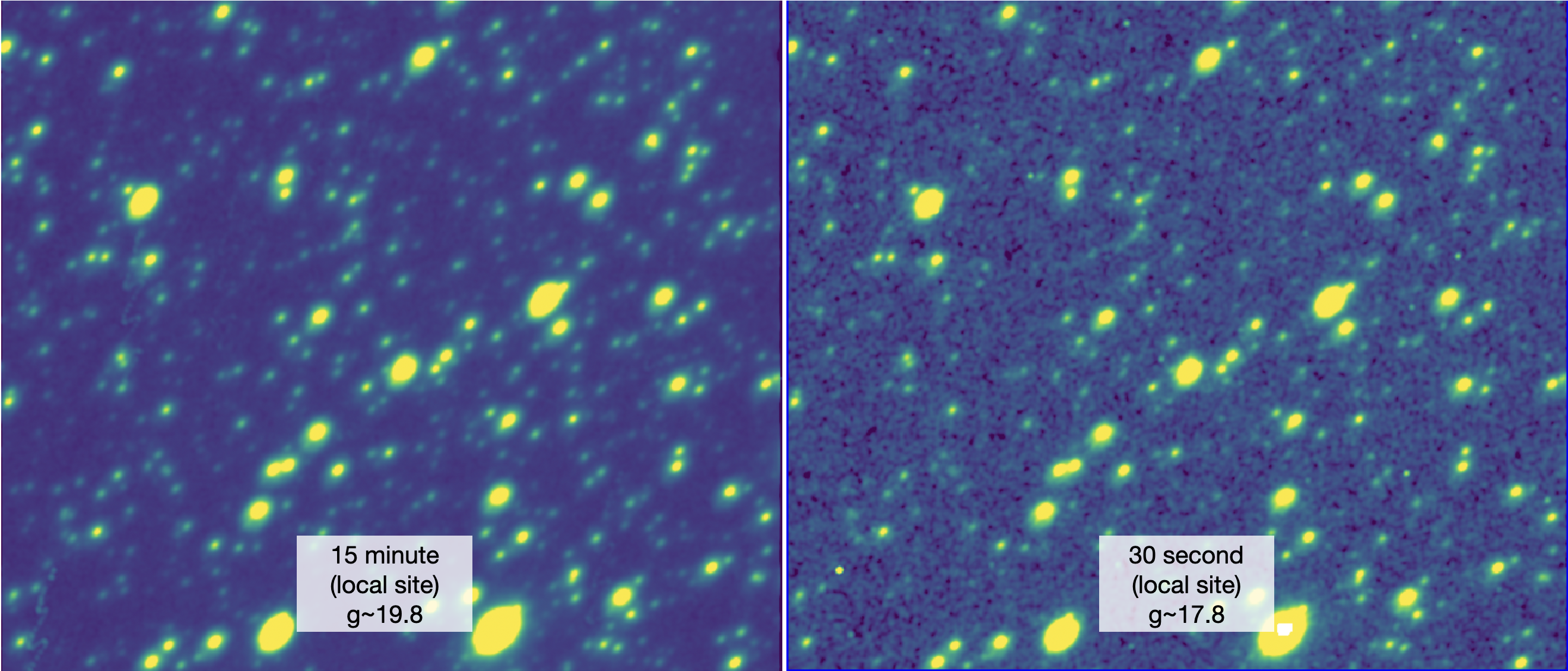}
    \end{tabular}
    \end{center}
    \caption{\label{fig:ratchet_coadd} A 15-minute coadd of data from a
    Celestron RASA8 node of A2TD. Full-frame images from each sensor are
    reprojected into HEALPix segments, which are then used for
    further analysis tasks (photometry, image subtraction, and coaddition.)}
 \end{figure}

Because the full-resolution image tiles are pre-aligned to a fixed HEALPix grid,
coaddition is greatly simplified. Coaddition on Argus sky tiles is optimized for
point-source detection using a per-image matched filter, which is the
statistically optimal method for background-limited
images.~\cite{2017ApJ...836..188Z} Figure~\ref{fig:ratchet_coadd} shows a deep
coadd of a set of 30$\times$ 30-second exposures, made using the algorithm
described in Ref.~\citenum{2017ApJ...836..188Z}. These images were collected
from a site local to our UNC-Chapel Hill lab, so the achievable depth is not
representative of median performance at a dark-sky site; however, the 5-$\sigma$
limiting magnitudes of $m_g=17.8$ at 30s and $m_g=19.8$ at 15 minutes are
reproducible from the same calculation we use to predict dark sky performance,
given typical sky brightness for a moderately-dark suburb ($m_{V}\sim19$ per sq.
arcsec). We reduced the images in Figure~\ref{fig:ratchet_coadd} using the
Argus-HDPS GPU/CPU pipeline, including image segmentation, reprojection, and
astrometry. 

\subsection{Transient Alerts}

\begin{figure}
    \begin{center}
    \begin{tabular}{c} 
    \includegraphics[width=1.0\columnwidth]{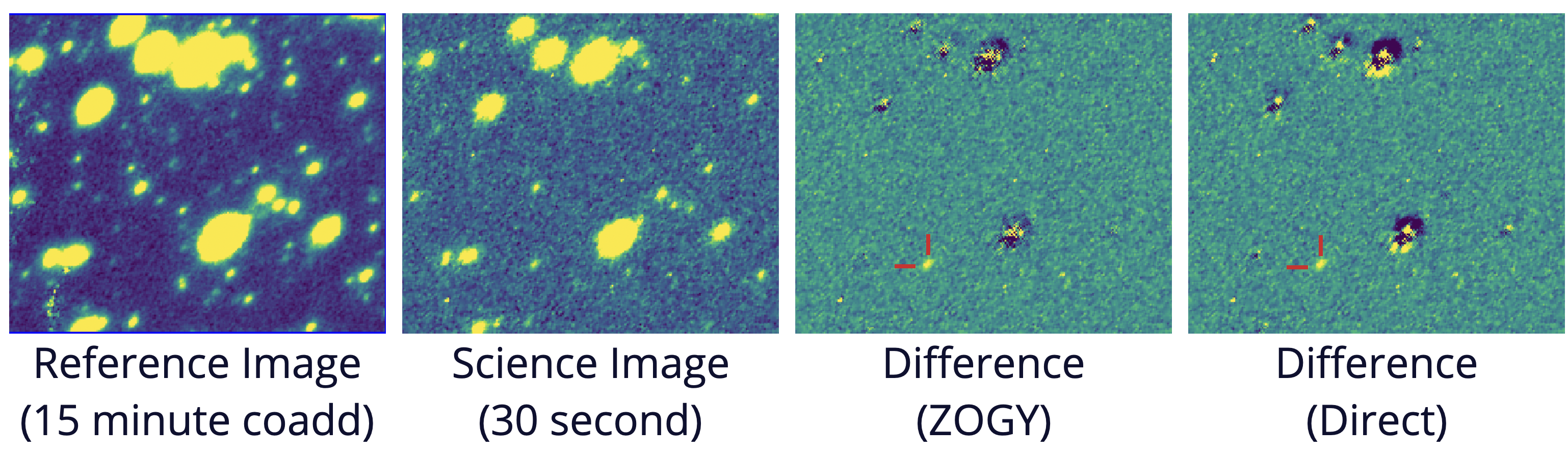}
    \end{tabular}
    \end{center}
    \caption{\label{fig:image_differencing} Per-tile image subtraction near an
    injected transient source with a 5-$\sigma$ peak significance, using
    both the high-speed direct subtraction algorithm of
    Ref.~\citenum{corbett_flashes}, and the ZOGY\cite{2016ApJ...830...27Z} algorithm. }
 \end{figure}

Argus Pathfinder will produce transient candidates at a rate of
approximately one million per night, and the full Argus Optical Array could meet
or exceed the alert rate of the Rubin Observatory. Transient detection is
performed using image subtraction in the reprojected HEALPix image segments,
leveraging long-term coadds as reference frames. We have implemented two
different algorithms for image subtraction in Argus-HDPS, and both will undergo
long-term on-sky evaluation with Argus Pathfinder; the ZOGY algorithm described
in Ref.~\citenum{2016ApJ...830...27Z} for image subtraction at base cadence and
in deep coadds at 15-minute cadence or greater, and the direct image subtraction
method we previously developed for the low-latency transient discovery pipeline for
the Evryscopes\cite{corbett_flashes}, for real-time operation in 1-second
cadence data.
Figure~\ref{fig:image_differencing} shows an image segment with a
simulated transient source, a reference image, and subtraction images made using
both subtraction algorithms. 

 In collaboration with the Arizona-NOIRLab Temporal Analysis and Response to
 Events System (ANTARES)\cite{2014SPIE.9149E..08S, 2021AJ....161..107M}, we are
 prototyping a public transient alert system for low-latency release of
 candidates from the Argus Optical Array. To maximize utility and ease of use
 for the community, we are adopting the evolving community standard of streaming
 serialized alert packets via Apache
 Kafka.\footnote{\url{https://kafka.apache.org/}}.

\subsection{Light Curves}

\begin{figure}
    \begin{center}
      \includegraphics[width=0.5\textwidth]{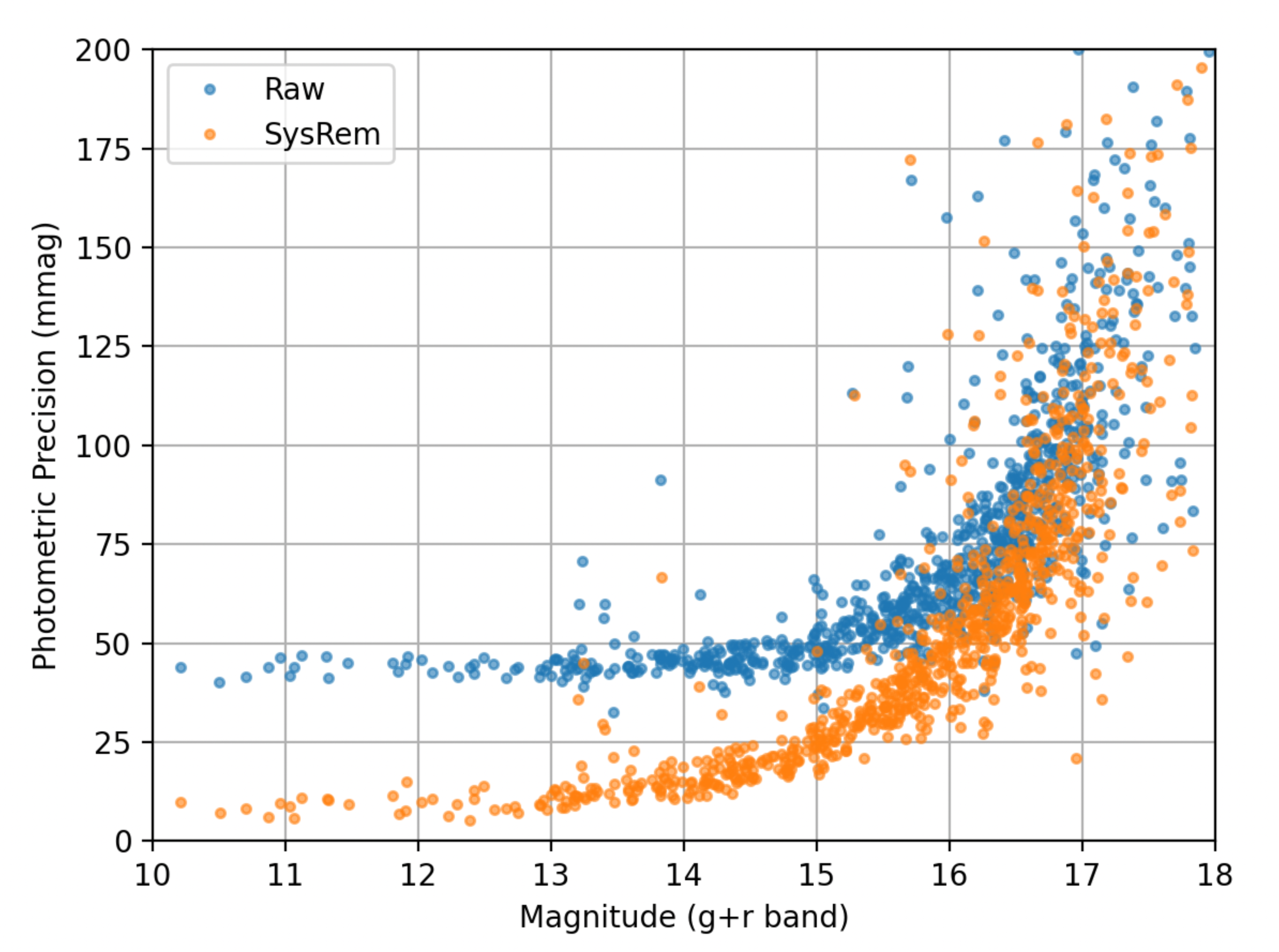}
      \caption{\label{fig:photometric_performance} Photometric RMS vs. Magnitude
      for 900 stars across a single 15-minute pointing at 30-second cadence.}
    \end{center}
 \end{figure}

Full evaluation of the expected precision of lightcurves from the Argus Optical
Array will be made using data from Argus Pathfinder; however,
initial results from A2TD nodes indicate that sub-percent performance may be
achievable, at least on short timescales.
Figure~\ref{fig:photometric_performance} shows the RMS vs magnitude for 900
stars during a single 15-minute ratchet at 30-second cadence, before and after
three iterations of detrending using the \textsc{sysrem}
algorithm\cite{tamuz_2005}, which achieves $\sim7$ mmag performance at the
bright end. 

\section{Pipeline Performance and Summary}\label{sec:summary}

We are developing Argus-HDPS, an integrated instrument control and data analysis
system for the Argus Optical Array, which uses asynchronous HTTP connections to
orchestrate observation and analysis tasks across multiple control servers
that pair optical/camera control with edge GPU computing capabilities for
minimal latency. For current prototyping stages, all analysis will be completed
on a single x86 server; scaling to a full 900-telescope system can be
accomplished linearly with the addition of 19 additional control servers. This
system enables a low-latency sharing of instrument state with an O(1000) camera
array, and allows for images to enter the analysis pipeline within 100 ms of camera
readout. We have demonstrated real-time image data product generation, using a
custom high-speed astrometric solver and GPU-based reprojection of sensor-plane
data to produce equal-area image segments, which are then coadded and subtracted
for transient detection at both single-image cadence and in deep coadds. Public
data release of imaging and transient alert data from the Argus Optical Array is
planned, and will be publicly prototyped after a commissioning period. 

Table~\ref{tbl:fullframe_timings} presents the average compute time for the
image segmentation and analysis pipeline stages, measured using a representative
36-core test server and an NVidia RTX 3090 Ti GPU. Source detection, astrometry,
and segmentation to the HEALPix grid are completed in an average of only 27
milliseconds on the GPU, which can be shared sequentially by many cameras to generate image
data products in real time, even at 1-second cadence. CPU-based image
subtraction is possible within cadence by parallelizing the direct subtraction
at the segment level; however, a GPU implementation is in development.

\begin{table}
    \centering
    \begin{tabular}{ll}
        \toprule
        \addlinespace[10pt]
        \multicolumn{2}{c}{GPU Timing}\\
        \midrule
        61-MPix image copy to GPU & 16 ms  \\
        Calibration & $<1$ ms  \\
        Median-filtered background map  & 1.7 ms  \\
        Source detection & 6.1 ms \\ 
        Image segmentation and resampling to HEALPix grid & 3.2 /ms \\ 

        \midrule
        \addlinespace[10pt]
        \multicolumn{2}{c}{CPU Timing (single-threaded)}\\
        \midrule
        Source de-duplication & 7.5 ms  \\
        Astrometric solution & 190 ms (95 ms with cached distortion terms)\\
        Image segments written to storage (modified FITS-format) & 475 ms\\
        Minipix stamping and low-resolution image generation & 300 ms (batch
        reduction)\\
        Direct image subtraction (1-second cadence) & 20 ms per tile\\
        ZOGY image subtraction (30-second+ cadence) & 1400 ms per tile\\
        \bottomrule
        \addlinespace[10pt]
    \end{tabular}

    \caption{\label{tbl:fullframe_timings} Average timing results for key
    processing steps in Argus-HDPS.}
\end{table}

\acknowledgments      
 
This paper was supported by NSF MSIP (AST-2034381) and CAREER (AST-1555175)
grants, and by the generosity of Eric and Wendy Schmidt by recommendation of the
Schmidt Futures program. This research and the construction of the Argus
prototypes in collaboration with the Be A Maker (BeAM)
network of makerspaces at UNC Chapel Hill and the UNC BeAM Design Center.

\clearpage
\bibliography{report}
\bibliographystyle{spiebib} 

\end{document}